\documentstyle[12pt]{article}
\begin {document}
\baselineskip 2.25pc

\begin{center}
{\Large Nonlocal electrodynamics in 2+1 dimensions from \\
	Chern-Simons theory}
\vskip 2pc
 Qiong-gui Lin\\
	Department of Physics, Zhongshan University, Guangzhou
        510275, P.R.China \footnote{Mailing Address}\\
        and\\
	China Center of Advanced Science and Technology (World
	Laboratory),\\
	P.O.Box 8730, Beijing 100080,P.R.China\\

\end{center}
\vfill
\centerline{\bf Abstract}

The theory of a spinor field interacting with a pure Chern-Simons
gauge field in 2+1 dimensions is quantized.  Dynamical and
nondynamical variables are separated in a gauge-independent way.
After the nondynamical variables are dropped, this theory reduces to a
pure spinor field theory with nonlocal interaction.  Several two-body
scattering processes are studied and the cross sections are obtained
in explicitly Lorentz invariant
forms.  
\vfill

\leftline {PACS number(s): 11.15.-q,  11.10.Lm}

\newpage

\section*{\large 1. Introduction}

Chern-Simons theories in 2+1 dimensions were widely
studied in recent years
by both condensed matter physicists and field theorists, due to its
potential applications to fractional quantum Hall effect and high
temperature superconductivity, as well as to its own significance in
gauge field theory.  To field theorists the Chern-Simons theory
without a Maxwell term [1] is of particular interest.  In this field
of investigation, many works have been devoted to classical soliton
solutions [2-6] and the problem of fractional spin and exotic
statistics [7,8].  In contrast, the quantum scattering processes have
not admitted enough attention.  In this circumstances, we are led to
consider the problem.

In a recent work [9], we dealt with the scattering problem of the
scalar Chern-Simons theory in some detail.  The theory was quantized
by Dirac's method [10].  Dynamical variables were separated from
nondynamical ones in a gauge-independent way by some appropriate
canonical transformations [11].  The quantized theory reduces to a
pure scalar theory with nonlocal interaction in the physical subspace
of the full Hilbert space.  Several two-body scattering processes were
studied there and the corresponding cross sections were obtained to
the lowest order in the coupling constant.

In this paper we extend the previous work [9] to the spinor Chern-
Simons theory.  We quantize it in the next section by the
Faddeev-Jackiw approach to quantization [12].  Though a complete
treatment in Dirac's manner can be given as in Ref.[9], the former
approach is much simpler in this case.  After the Hamiltonian
reduction is carried out, we perform an appropriate canonical
transformation which enables us to separate dynamical variables from
nondynamical ones in a gauge-independent way.  Then the nondynamical
variables can be dropped, and the theory reduces to one that involves
only the spinor field with nonlocal interaction.  Quantization of the
theory becomes thereupon straightforward.  In Sec. 3, we calculate the
cross sections of several two-body scattering processes.  This is
carried out within the framework of perturbation theory, and only the
lowest-order contribution is calculated.  Higher order correction
remains to be further investigated.  Though the theory seems simpler
than ordinary QED in its original form, the calculation is much more
complicated than that in QED.  Moreover, to bring the cross sections
into explicitly Lorentz invariant forms, one has to make out some
Lorentz invariants in the scattering processes which are not
manifestly invariant.  

\section*{\large 2. Quantization and perturbation expansion}

     Let us begin with the (2+1)-dimensional Lagrangian density
$$ {\cal L}=i\bar\psi\gamma^\mu D_\mu\psi -m\bar\psi\psi +
{\kappa\over 2}\epsilon^{\lambda\mu\nu} A_\lambda\partial_\mu A_\nu
\eqno(2.1)$$
where $ D_\mu=\partial_\mu+ieA_\mu$, $\epsilon^{012}=1$, $\psi$
is a two-component massive spinor field with mass $m$, $A_\mu$   is
the Chern-Simons gauge field, $e$ is the coupling constant. It will
be seen
below, however, that the physically relevant coupling strength is
governed by $e^2/\kappa$. In (2.1) $\gamma^\mu$
are the Dirac matrices satisfying
$$\{ \gamma^\mu, \gamma^\nu\}=2g^{\mu\nu}.\eqno(2.2)$$
We use the metric $g^{\mu\nu}={\rm diag}(1, -1, -1)$.
The algabra (2.2) can be realized by the Pauli matrices:
$$ \gamma^0=\sigma^3, \gamma^1=i\sigma^1, \gamma^2=i\sigma^2.
\eqno(2.3a)$$
In the representation (2.3a) we have
$$\gamma^\mu \gamma^\nu=g^{\mu\nu}-i\epsilon^{\mu\nu\lambda}
\gamma_\lambda.\eqno(2.3b)$$
This holds for all representations equivalent to (2.3a). In another
representation, say,
$$ \gamma^0=-\sigma^3, \gamma^1=i\sigma^1, \gamma^2=i\sigma^2
\eqno(2.4a)$$
which is not equivalent to (2.3a), we have
$$\gamma^\mu \gamma^\nu=g^{\mu\nu}+i\epsilon^{\mu\nu\lambda}
\gamma_\lambda\eqno(2.4b)$$
instead of (2.3b).  Again this is valid for all representations
equivalent to (2.4a).  The difference between (2.3b) and (2.4b) 
would not affect any physical result.

The theory described by (2.1) can be quantized by Dirac's method.
However, a complete treatment in this manner is rather lengthy.  Since
the Lagrangian density is linear in the ``velocities'', we prefer the
Faddeev-Jackiw approach to quantization, which is much simpler in
this case.  The Lagrangian associated with (2.1) can be written as
$$L=\int d{\bf
x}\;\left[i\bar\psi\gamma^0\dot\psi+\frac\kappa2\epsilon^{ij} A_j\dot
A_i+ i\bar\psi\gamma^i D_i\psi-m\bar\psi\psi- A_0
(e\bar\psi\gamma^0\psi-\kappa\epsilon^{ij}\partial_i A_j)\right].
\eqno(2.5)$$
Variation of the action with respect to $A_0$ gives rise
to the constraint
$$\xi\equiv\rho+\kappa B\approx 0\eqno(2.6)$$
where $B=\epsilon^{ij}\partial_i A^j $ is the magnetic field and
$\rho=e\bar\psi\gamma^0\psi$ is the charge density.  Thus $A_0$ is not
a dynamical variable.  It just plays the role of a Lagrange
multiplier.  Variations of the action with respect to $\psi$,
$\bar\psi$ and $A_i$ gives their equations of motion.
We rewrite (2.5) in the form
$$L=\int d{\bf x}\;\left[i\bar\psi\gamma^0\dot\psi+\frac\kappa2
\epsilon^{ij} A_j\dot A_i\right]-H_T,\eqno(2.7a)$$
$$H_T=\int d{\bf x}\;[-i\bar\psi\gamma^i D_i\psi
+m\bar\psi\psi+ A_0\xi].  \eqno(2.7b)$$
It can be easily
verified that the Lagrange equation of motion for any canonical
variable $\Phi$($\psi$,$\bar\psi$ or $A_i$) can take the Hamilton form
$$\dot\Phi=\{\Phi, H_T\}^*\eqno(2.8)$$
if the following algebra is posited.
$$\{\psi({\bf x}), \bar\psi({\bf y})\}^*=-i\gamma^0
\delta({\bf x} -{\bf y}),\eqno(2.9a)$$
$$\{A_i({\bf x}),A_j({\bf y})\}^*=\frac 1\kappa\epsilon_{ij}
\delta({\bf x} -{\bf y}),\eqno(2.9b)$$
where the time $t$ common to all variables is suppressed.(It will
be suppressed throughout the paper for simplicity
in notations.) Here the ``*'' indicates that the brackets are not the
ordinary Poisson ones.  Indeed they are nothing less than the Dirac
brackets one could find in the Dirac approach to quantization.

     From (2.9) we see that ($A_1$, $\kappa A_2$) is a canonical
pair. This is not convenient for further treatment. Thus we
decompose $A_i$   into its longitudinal and transverse components as
$$A_i=A_i^L+A_i^T\equiv\partial_i\omega+\epsilon_{ij}
\partial_j\eta.\eqno(2.10)$$
It is easy to show that $\nabla^2\eta=B$,and the only nonvanishing
brackets among $\omega({\bf x})$ and $B({\bf x})$ is
$$\{ \omega({\bf x}),\kappa B({\bf y}) \}^*= \delta({\bf x} -{\bf y}).
\eqno(2.11)$$
Substituting (2.10) into (2.7) and dropping some surface terms we get
$$L=\int d{\bf x}\;[i\bar\psi\gamma^0\dot\psi+\kappa B\dot\omega]-H_T
\eqno(2.12)$$
where $H_T$ is still given by (2.7b) but where $A_i$ is expressed in
$\omega$  and $B$ by using the relation
$$A_i^T({\bf x})=\int d{\bf y}\;\epsilon_{ij}\partial_j G
({\bf x}-{\bf y})B({\bf y}).\eqno(2.13)$$
In this equation the partial derivative $\partial_j$ is with respect
to {\bf x} and $G({\bf x})$ is the Green function
$$G({\bf x})={1 \over 2\pi}\ln |{\bf x}|\eqno(2.14a)$$
satisfying
$$\nabla^2G({\bf x})=\delta({\bf x}).\eqno(2.14b)$$
The canonical pair ($A_1$,$\kappa A_2$) is thereby replaced by
($\omega$, $\kappa B$).

In order to eliminate nondynmical variables, we now introduce the
canonical transformation [11]
$$\psi=\exp(-ie\omega)\Psi,\bar\psi=\exp(ie\omega)\bar\Psi.
\eqno(2.15)$$
Under this transformation, (2.12) is changed to
$$L=\int d{\bf x}\;[i\bar\Psi\gamma^0\dot\Psi+
\xi\dot \omega]-H_T,\eqno(2.16a)$$
$$H_T=\int d{\bf x}\;[-i\bar\Psi\gamma^i D_i^T\Psi
+m\bar\Psi\Psi+ \lambda\xi],  \eqno(2.16b)$$
where $D_i^T=\partial_i+ieA_i^T$,and
$$A_i^T({\bf x})=-\frac 1\kappa\int d{\bf y}\;\epsilon_{ij}
\partial_j G({\bf x}-{\bf y})\rho({\bf y}),\eqno(2.17a)$$
$$\rho=e\bar\Psi\gamma^0\Psi.\eqno(2.17b)$$
In obtaining the results we have redefined the Lagrange multiplier
$A_0$ and denoted it by $\lambda$.  The brackets become
$$\{\Psi({\bf x}), \bar\Psi({\bf y})\}^*=-i\gamma^0
\delta({\bf x} -{\bf y}),\eqno(2.18)$$
$$\{\omega({\bf x}),\xi({\bf y})\}^*=\delta({\bf x} -{\bf y}).
\eqno(2.19)$$
Obviously, the canonical pair ($\omega$, $\kappa B$) has been further
replcaed by ($\omega$, $\xi$). It is remarkable that the constraint
$\xi$ becomes a canonical momentum. Because of (2.6), the canonical
pair ($\omega$, $\xi$) are not dynamical variables
and thus can be simply dropped.  Then (2.16) becomes
$$L=\int d{\bf x}\;i\bar\Psi\gamma^0\dot\Psi-H_T,\eqno(2.20a)$$
$$H_T=\int d{\bf x}\;[-i\bar\Psi\gamma^i D_i^T\Psi
+m\bar\Psi\Psi].\eqno(2.20b)$$
Hence we arrived at a theory described by (2.20), (2.17) and (2.18).
It involves only the spinor field $\Psi$ with nonlocal interaction.
Since $\{\Psi({\bf x}), \xi({\bf y})\}^*=0$, the spinor field $\Psi$
is gauge invariant and thus seems more
``physical'' than the original $\psi$.

Now it is easy to accomplish the transition from classical to quantum
theory.  First the bracket (2.18) is promoted to an anticommutation
relation
$$\{\Psi({\bf x}), \bar\Psi({\bf y})\}=\gamma^0
\delta({\bf x} -{\bf y}).\eqno(2.21)$$
Then the equation of motion is replaced by
$$i\dot\Psi=[\Psi, H_T]\eqno(2.22)$$
where the commutator is evaluated by using (2.21), and
$$H_T=\int d{\bf x}\;:[-i\bar\Psi\gamma^i D_i^T\Psi
+m\bar\Psi\Psi]: \eqno(2.23)$$
where $A_i^T$ is given by (2.17a) with
$$\rho=e:\bar\Psi\gamma^0\Psi:.\eqno(2.24)$$
Since $H_T$ involves ordering ambiguity, we have adopted the
normal-ordering prescription denoted by colons, which will become
clear in the interaction picture (see below).  The normal-ordering of
$\rho$ just serves to remove a zero-point charge.

Similar to ordinary QED, we decompose $H_T$ into two parts, the free
Hamiltonian $H_0$ and the interacting one $H$:
$$ H_T=H_0+H,\eqno(2.25a)$$
$$H_0=\int d{\bf x}\;:[-i\bar\Psi\gamma^i \partial_i\Psi
+m\bar\Psi\Psi]:, \eqno(2.25b)$$
$$H=e\int d{\bf x}\;:\bar\Psi\gamma^i A_i^T\Psi:,\eqno(2.25c)$$
and go to the interaction picture where $\Psi$ etc.  are transformed
to $\Psi_I$ etc..  In the following we work in the interaction picture
but omit the subscript $I$.  In this picture the anticommutation
relation (2.21) remains unchanged.  The field operator $\Psi$
obeys the equation
$$i\dot\Psi=[\Psi, H_0]\eqno(2.26)$$
while the evolution of a physical state $|P\rangle$ is governed by
$$ i{\partial \over\partial t}|P\rangle=H|P\rangle.\eqno(2.27)$$
These are well known in quantum field theory.  However, in 2+1
dimensions there is something different and must be described in some
detail.  On account of (2.26) and (2.25b) we have for $\Psi$ the
equation
$$ (i\gamma^\mu\partial_\mu-m)\Psi=0.\eqno(2.28)$$
This is nothing but the Dirac equation in 2+1 dimensions, as it should
be.  The field operator $\Psi$ can thus be expanded as
$$\Psi(x)={1\over\sqrt V}\sum_{\bf k}(a_{\bf k}u_{\bf k}e^
{-ik\cdot x}+b_{\bf k}^\dagger v_{\bf k}e^{ik\cdot x})\eqno(2.29)$$
where $V$ is a two-dimensional normalization volume, $k_0=\sqrt
{{\bf k}^2+m^2}$ is positive, and $u_{\bf k}$, $v_{\bf k}$
satisfy the following equations.
$$ (\gamma^\mu k_\mu-m)u_{\bf k}=0,\quad
 (\gamma^\mu k_\mu+m)v_{\bf k}=0.\eqno(2.30)$$
The solutions $u_{\bf k}$, $v_{\bf k}$ are normalized as
$$u_{\bf k}^\dagger u_{\bf k}=1,\quad v_{\bf k}^\dagger v_{\bf k}=1.
\eqno(2.31)$$
It should be remarked that there is no spin index for the spinors
$u_{\bf k}$ and $v_{\bf k}$ in 2+1 dimensions, since for a given
{\bf k} there is only one linearly independent solution to any
one of (2.30).
The spin operator with regard to Eq.(2.28) can be shown to be
$$J=\frac i4\epsilon^{ij}\gamma^i\gamma^j.\eqno(2.32)$$
In the representation (2.3) or (2.4), we have
$$ J=\pm\frac12\gamma^0\eqno(2.33)$$
where the upper(lower) sign corresponds to (2.3)((2.4)).
On the other hand, (2.30) reduces to
$$\gamma^0u_{\bf k}=u_{\bf k},\quad \gamma^0v_{\bf k}=-v_{\bf k}
\eqno(2.34)$$
when ${\bf k}=0$.  Therefore when ${\bf k}=0$ $u_{\bf k}$ has spin
1/2 or $-1/2$ according as the representation (2.3) or (2.4) is
employed, while $v_{\bf k}$ has the opposite spin to $u_{\bf k}$.
The discussion also holds for those representations
equivalent to (2.3) or (2.4).  From (2.30) one can derive the
orthogonal relations
$$\bar u_{\bf k}v_{\bf k}=0, \quad \bar v_{\bf k}u_{\bf k}=0.\eqno(2.35)$$
Eqs.(2.31) and (2.35) enables us to find from (2.21) the
nonvanishing anticommutators among $a_{\bf k}$, $b_{\bf k}$ etc.:
$$\{a_{\bf k},a_{\bf l}^\dagger\}=\delta_{\bf kl},\quad
\{b_{\bf k},b_{\bf l}^\dagger\}=\delta_{\bf kl}.\eqno(2.36)$$
Similar to ordinary QED, it can be shown that
$$u_{\bf k}\bar u_{\bf k}={\gamma\cdot k+m \over 2k_0},\quad
v_{\bf k}\bar v_{\bf k}={\gamma\cdot k-m \over 2k_0},\eqno(2.37)$$
which will be useful in the next section.  Again we emphasize that
there is no spin summation in (2.37), which is different from the
case in 3+1 dimensions.

As in ordinary QED, $a_{\bf k}^\dagger$ is the creation operator of a
particle with momentum {\bf k}, energy $k_0$, and charge $e$, while
$a_{\bf k}$ is the corresponding annihilation operator.  Similarly,
$b_{\bf k}^\dagger$ creates an antiparticle with momentum {\bf k},
energy $k_0$, and charge $-e$, while $b_{\bf k}$ annihilates it.

The transition amplitude for the system from an initial state
$|i\rangle$ to a final one $|f\rangle$ is given by
$$ A_{fi}=\langle f|S|i\rangle \eqno(2.38)$$
where $S$ is the scattering operator given by
$$ S=1+\sum_{n=1}^\infty S^{(n)},\eqno(2.39a)$$
$$ S^{(n)}={(-i)^n \over n!}\int_{-\infty}^{+\infty} dt_1\cdots
dt_n\;T[H(t_1)\cdots H(t_n)].\eqno(2.39b)$$
It can be seen from (2.25c) and (2.17) that $S^{(n)}$ is of the order
$(e^2/\kappa)^n$, thus the ratio $e^2/\kappa$ is similar to the
electromagnetic coupling constant in QED, governing the strength of
coupling. For two-body scattering, the lowest-order contribution
to $S$ comes from $S^{(1)}$:
$$S^{(1)}=i{e^2\over \kappa}\int dt\;d{\bf x}\;d{\bf y}\;\epsilon^
{ij}\partial_j G({\bf x}-{\bf y}):\bar\Psi({\bf x})\gamma^i\Psi
({\bf x})\bar\Psi({\bf y})\gamma^0\Psi({\bf y}):.\eqno(2.40)$$

     With the preparations of this section, we are now equipped to
investigate two-body scattering processes in detail.

\section*{\large 3. Cross sections and Lorentz invariance}

For two-body scattering , the transition amplitude (2.38) can be
written in the form
$$ A_{fi}=(2\pi)^3 V^{-2}\delta(p+q-k-l)R(p,q,k,l)\eqno(3.1)$$
where $k$, $l$ are the initial three-momenta of the two particles and
$p$, $q$ are final ones, the function $R(p,q,k,l)$ depends on the
particular process.  By the method similar to that used in QED[13],
the cross section (in two spatial dimensions it may be more
appropriately called the cross width) can be found to be
$$\sigma={1 \over 2\pi|{\bf k}/k_0-{\bf l}/l_0|}\int d\theta\,
{p_0|R|^2\over \partial(p_0+q_0)/\partial p_0}\eqno(3.2)$$
where $\theta$ is the angle between {\bf p} and the incident direction
which is defined as the direction of the momentum of the incident
particle in the laboratory system.  This is subject to the condition
$$ {\bf p}+{\bf q}={\bf k}+{\bf l}\eqno(3.3a)$$
by which $q_0$ depends on $p_0$ (or $|{\bf p}|$) and $\theta$.
In evaluating the partial derivative $\partial q_0/\partial p_0$,
$\theta$ is treated as a constant.  Finally the result (3.2)
is further subject to the additional condition
$$ p_0+q_0=k_0+l_0 \eqno(3.3b)$$
which, together with (3.3a), exhibit conservation of the total
momentum and energy.  The integration bounds of $\theta$ in (3.2)
depends on the frame of reference where the calculation is carried
out.  It also depends on the particular process dealt with.

The Mandelstam invariants defined as
$$ s=(k+l)^2=2k\cdot l+2m^2,\eqno(3.4a)$$
$$ t=(p-k)^2=-2p\cdot k +2m^2,\eqno(3.4b)$$
$$ u=(p-l)^2=-2p\cdot l +2m^2 \eqno(3.4c)$$
will be employed below.  The invariant $t$ should not be confused
with the time.  On account of (3.3) we have
$$ s+t+u=4m^2.\eqno(3.5)$$
The range of values for a physical scattering process is
$$ s>4m^2, \quad t\le 0, \quad u\le 0.\eqno(3.6)$$

     Before going into the details of any particular process, we
would like to recast the formula (3.2) into a form convenient for
writing it in a manifestly Lorentz invariant manner.  First of all
it should be emphasized that invariance of the cross section is
justified only in those frames of reference where ${\bf k}\parallel
{\bf l}$.({\bf l} points at the same or opposite direction of
{\bf k}.)  In the following such frames of reference are
called parallel systems, and the Lorentz transformations among these
systems are called parallel boosts.  In these systems, 
one can show that
$$\sigma={1 \over\pi s\sqrt{s-4m^2}}\int dt\,{p_0q_0k_0l_0|R|^2\over
\sqrt{tu}}.\eqno(3.7)$$
For a Lorntz invariant theory, one expects on the basis of (3.7) that
$p_0q_0k_0l_0|R|^2$ might be written in a Lorentz invariant form.
When this is done, (3.7) gives the explicitly Lorentz invariant
expression for the cross section.  Of course, the integration bounds
should also be Lorentz scalars.

We are now ready to calculate the cross sections for various
scattering processes.  For particle-particle scattering,
the initial and final states are
$$|i\rangle=a_{\bf k}^\dagger a_{\bf l}^\dagger|0\rangle,\quad
|f\rangle=a_{\bf p}^\dagger a_{\bf q}^\dagger|0\rangle \eqno(3.8)$$
where $|0\rangle$ is the vacuum state defined by
$$a_{\bf k}|0\rangle=0,\quad b_{\bf k}|0\rangle=0,\quad \forall {\bf k}.
\eqno(3.9)$$
The transition amplitude is given by (3.1) where $R$ can be shown to
be
$$R(p,q,k,l)=\sum_{i=1}^4 R_i(p,q,k,l)\eqno(3.10a)$$
where
$$R_1(p,q,k,l)={e^2\over \kappa}{\epsilon^{ij}(p-k)^i\over({\bf p-k})
^2}\bar u_{\bf q}\gamma^j u_{\bf l}\bar u_{\bf p}\gamma^0 u_{\bf k},
\eqno(3.10b)$$
$$R_2(p,q,k,l)=-R_1(p,q,l,k),\eqno(3.10c)$$
$$R_3(p,q,k,l)=-R_1(q,p,k,l),\eqno(3.10d)$$
$$R_4(p,q,k,l)=R_1(q,p,l,k).\eqno(3.10e)$$
Our main job is to calculate $|R|^2$. This essentially reduces to the
evaluation of the traces of various products of the $\gamma$ matrices,
due to the relation (2.37).  The evaluation of such traces is
similar to those in 3+1 dimensions when an even number of $\gamma$'s
is involved.
The product of an odd number of $\gamma$ matrices in 2+1 dimensions has
in general nonvanishing trace, however.  For example,
$${\rm tr}(\gamma^\mu\gamma^\nu\gamma^\lambda)=\mp 2i\epsilon^
{\mu\nu\lambda}\eqno(3.11)$$
where the upper(lower) sign corresponds to the representatin (2.3)
((2.4)) or its equivalent representations.  The different signs in
(3.11) do not influence the final result.  The trace of the product of
five $\gamma$'s may also be of use in the calculation, but not
mandatory. What is important is that the trace of the product of an
odd number $\gamma$'s is imaginary while that of an even number ones
is real.  This may simplify some calculations.  The following
relations are also useful.
$$\gamma^i\gamma\cdot k\gamma^i=2\gamma^0 k^0,\quad
\gamma^0\gamma\cdot k\gamma^0=\gamma\cdot\bar k\eqno(3.12)$$
where $k$ is an arbitrary Lorentz vector and $\bar k$ is defined as
$$\bar k^\mu=(k^0, -{\bf k})\eqno(3.13)$$
such that
$$\bar p\cdot k=p\cdot \bar k=p^0k^0+{\bf p\cdot k}\eqno(3.14)$$
where $p$ is another Lorentz vector.  Of course $\bar k$ is not a
Lorentz vector and $p\cdot\bar k$  is not a Lorentz scalar.  Since
the evaluation of $|R|^2$ is
very lengthy we introduce some notations defined as follows.
$$P_i=R_i^*R_i ~({\rm no~ summation~ over}~ i), \quad i=1,2,3,4,
\eqno(3.15a)$$
$$P_{ij}=R_i^*R_j+R_iR_j^*, \quad 1\le i<j\le 4.\eqno(3.15b)$$
$${\bf a=p-k,\quad b=p-l.}\eqno(3.16)$$
With the above preparations, we write down the following results
obtained from straightforward but lengthy calculations.
$$P_{12}+P_{34}+P_{13}+P_{24}=-{e^4\over\kappa^2}{s-4m^2\over
2p_0q_0k_0l_0},\eqno(3.17)$$
$$P_1+P_4+P_{14}={e^4\over\kappa^2}{m^2\over p_0q_0k_0l_0}+
{e^4\over\kappa^2}{1\over 4p_0q_0k_0l_0}
\left[{(q_0+l_0){\bf k\times p}-(p_0+k_0){\bf l\times p}\over
{\bf a}^2}\right]^2\eqno(3.18)$$
where ${\bf k\times p}=\epsilon^{ij}k^ip^j$ etc..
The next step is to bring the quantity in the above square bracket
into a manifestly Lorentz invariant form, and simplify the above
equation at the same time. This is not very easy since neither the
denominator nor the numerator is Lorentz invariant. To our
knowledge similar case was not encountered in ordinary QED. So we
will describe it in some detail.
Obviously, ${\bf k\times p}=k_\parallel p_\perp$ where $k_\parallel$
is the component of {\bf k} in the incident direction($k_\parallel=
\pm|{\bf k}|$) and $p_\perp$ that of {\bf p} in the perpendicular
direction.  So
$${(q_0+l_0){\bf k\times p}-(p_0+k_0){\bf l\times p}\over
{\bf a}^2}=
{2(l_0k_\parallel-k_0l_\parallel)+(k_0-p_0)(k_\parallel+l_\parallel)
\over {\bf a}^2}
p_\perp.\eqno(3.19)$$
$p_\perp$ is obviously invariant under parallel boosts.
It is easy  to show that $l_0k_\parallel-k_0l_\parallel$ 
is also Lorentz invariant in the parallel systems.
Of particular importance here is that
the same thing can be proved for $(k_\parallel
+l_\parallel)/(k_0-p_0)$.
We choose ${\bf l}=0$ in the laboratory system
without loss of generality.  Then it can be easily verified that
$$p_\perp^2={tu\over s-4m^2}\eqno(3.20a)$$
$$l_0k_\parallel-k_0l_\parallel={\sqrt{s(s-4m^2)}\over 2},\eqno
(3.20b)$$
$${k_\parallel+l_\parallel\over k_0-p_0}=-{\sqrt{s(s-4m^2)}\over
t}.\eqno(3.20c)$$
Indeed, the first two relations have been used in obtaining (3.7).
If we choose ${\bf k}=0$ in the laboratory system, the last two relations
will be modified by a minus sign.  This, however, does not affect the
final result.  
Collecting these results we obtain
$$P_1+P_4+P_{14}={e^4\over\kappa^2}{m^2\over p_0q_0k_0l_0}+
{e^4\over\kappa^2}{1\over 4p_0q_0k_0l_0}{su\over t}.\eqno(3.21)$$
By interchanging $k$ and $l$ and thus $t$ and $u$ in the result
(3.21) we obtain
$$P_2+P_3+P_{23}={e^4\over\kappa^2}{m^2\over p_0q_0k_0l_0}+
{e^4\over\kappa^2}{1\over 4p_0q_0k_0l_0}{st\over u}.\eqno(3.22)$$
These two equations, together with (3.17), give the final result of
$|R|^2$:
$$|R|^2=\sum_{i=1}^4 P_i+\sum_{1\le i<j\le 4}P_{ij}=
{e^4\over\kappa^2}{4m^2\over p_0q_0k_0l_0}+
{e^4\over\kappa^2}{1\over 4p_0q_0k_0l_0}{s(t-u)^2\over tu}.
\eqno(3.23)$$
In the center-of-mass system, it is easy to show that
$$|R|^2={e^4\over\kappa^2}{4(m^2+k_0^2\cot^2\theta)\over k_0^4}.
\eqno(3.24)$$
The cross section, expressed in the form (3.2), turns out to be
$$\sigma_{pp}={e^4\over\kappa^2}{1\over 2\pi|{\bf k}|k_0^2}\int_0
^\pi d\theta\,(m^2+k_0^2\cot^2\theta).\eqno(3.25)$$
The integration of $\theta$ is performed only for $\theta>0$ since
the two particles are identical.  Taking this into account and
employing (3.7) we obtain the explicitly Lorentz invariant
expression for the cross section:
$$\sigma_{pp}={e^4\over\kappa^2}{1\over 4\pi s\sqrt{s-4m^2}}\int^0
_{4m^2-s} dt\,{s(t-u)^2+16m^2tu\over (tu)^{3/2}}.\eqno(3.26)$$
From (3.26) one can easily acquire the expression of $\sigma_{pp}$
in the laboratory system or other systems.

For antiparticle-antiparticle scattering, the result is completely the
same as that of particle-particle scattering.  The final process we
are going to study is the scattering between a particle and an
antiparticle.  In this case
$$|i\rangle=a_{\bf k}^\dagger b_{\bf l}^\dagger|0\rangle,\quad
|f\rangle=a_{\bf p}^\dagger b_{\bf q}^\dagger|0\rangle. \eqno(3.27)$$
The transition amplitude is given by (3.1) where R is replaced by
$\tilde R$, and
$$\tilde R(p,q,k,l)=\sum_{i=1}^4\tilde R_i(p,q,k,l)\eqno(3.28a)$$
where
$$\tilde R_1(p,q,k,l)={e^2\over\kappa}{\epsilon^{ij}a^i\over{\bf a}
^2}\bar u_{\bf p}\gamma^j u_{\bf k}\bar v_{\bf l}\gamma^0 v_{\bf q},
\eqno(3.28b)$$
$$\tilde R_2(p,q,k,l)=-{e^2\over\kappa}{\epsilon^{ij}a^i\over{\bf a}
^2}\bar v_{\bf l}\gamma^j v_{\bf q}\bar u_{\bf p}\gamma^0 u_{\bf k},
\eqno(3.28c)$$
$$\tilde R_3(p,q,k,l)={e^2\over\kappa}{\epsilon^{ij}c^i\over{\bf c}
^2}\bar v_{\bf l}\gamma^j u_{\bf k}\bar u_{\bf p}\gamma^0 v_{\bf q},
\eqno(3.28d)$$
$$\tilde R_4 (p,q,k,l)=-{e^2\over\kappa}{\epsilon^{ij}c^i\over{\bf c}
^2}\bar u_{\bf p}\gamma^j v_{\bf q}\bar v_{\bf l}\gamma^0 u_{\bf k}
\eqno(3.28e)$$
where {\bf a} is defined in (3.16) and
$${\bf c=k+l}.\eqno(3.29)$$
Since the evaluation of $|\tilde R|^2$ is also tedious, we define
the quantities $\tilde P_i$, $\tilde P_{ij}$ in a similar way as
(3.15).  It is not difficult to find that
$$\tilde P_1=P_4,\quad \tilde P_2=P_1, \quad \tilde P_{12}=P_{14}.
\eqno(3.30)$$
Thus
$$\tilde P_1+\tilde P_2+\tilde P_{12}=P_1+P_4+P_{14}
={e^4\over\kappa^2}{m^2\over p_0q_0k_0l_0}+
{e^4\over\kappa^2}{1\over 4p_0q_0k_0l_0}{su\over t}.\eqno(3.31)$$
In the parallel systems one can show that
$$\tilde P_3+\tilde P_4+\tilde P_{34}
={e^4\over\kappa^2}{m^2\over p_0q_0k_0l_0}+
{e^4\over\kappa^2}{1\over 4p_0q_0k_0l_0}{tu\over s}.\eqno(3.32)$$
The calculation is similar to that carried out above for
particle-particle scattering, and we have realized the fact
that $(k_0-l_0)/(k_\parallel+l_\parallel)$
is invariant under parallel boosts and employed the relation
(as before, we choose ${\bf l}=0$ in the laboratory system)
$${k_0-l_0\over k_\parallel+l_\parallel}=\sqrt{s-4m^2\over s}.
\eqno(3.33)$$
After lengthy but straightforward calculations it turns out that
$$\tilde P_{13}+\tilde P_{14}+\tilde P_{23}+\tilde P_{24}=
{e^4\over\kappa^2}{4m^2-u\over 2p_0q_0k_0l_0}.\eqno(3.34)$$
The evaluation of $|\tilde R|^2$ is accomplished by gathering the
above results and the final result reads
$$|\tilde R|^2=\sum_{i=1}^4 \tilde P_i+\sum_{1\le i<j\le 4}\tilde
P_{ij}={e^4\over\kappa^2}{4m^2\over p_0q_0k_0l_0}+
{e^4\over\kappa^2}{1\over 4p_0q_0k_0l_0}{u(t-s)^2\over ts}.
\eqno(3.35)$$
This can be obtained by interchanging $p$ and $-k$ and thus $u$ and
$s$ in the previous result (3.23) for particle-particle scattering.
However, there seems not sufficient reason to get (3.35) by such an
interchange at the beginning.  In the center-of-mass system,
it is easy to show that
$$|\tilde R|^2={e^4\over\kappa^2}{4m^2\over k_0^4}+{e^4\over\kappa^2}
{1\over k_0^2}\cot^2{\theta\over 2}\left(1+{{\bf k}^2\over k_0^2}
\sin^2{\theta\over 2}\right)^2.\eqno(3.36)$$
The cross section expressed by (3.2) turns out to be
$$\sigma_{pa}={e^4\over\kappa^2}{1\over8\pi|{\bf k}|}\int_{-\pi}^\pi
d\theta\,\left[{4m^2\over k_0^2}+\cot^2{\theta\over2}\left(1+
{{\bf k}^2\over k_0^2}\sin^2{\theta\over2}\right)^2\right].
\eqno(3.37)$$
Here the integration is performed over the full range of $\theta$
since the two particles are distinguishable.  The explicitly Lorentz
invariant form of $\sigma_{pa}$ is easily acquired by substituting
(3.35) into the expression (3.7):
$$\sigma_{pa}={e^4\over\kappa^2}{1\over 2\pi s^2\sqrt{s-4m^2}}\int^0
_{4m^2-s} dt\,{u(t-s)^2+16m^2ts\over t\sqrt{tu}}.\eqno(3.38)$$
From this general result the expression of $\sigma_{pa}$ in the
laboratory system can be easily obtained.

We have thus finished the calculation of the cross sections for the
several two-body scattering processes in this theory.  The calculation
seems much more complicated than that in ordinary QED, though the
theory seems simpler in its original form.  In order to put the cross
sections into manifestly Lorentz invariant forms, one has to work
out some relations such as (3.20c) and (3.33), which were not  
encountered in ordinary QED to our knowledge. The
calculation is carried out only to the lowest order in $e^2/\kappa$,
or at the tree level. Higher-order corrections remain to be further
investigated. It may be expected that the calculation of this would
be still more complicated than that presented here.
\vskip 1pc

This work was supported by the
Doctoral Programme Foundation of the National Education Commission
of China and by the National Natural Science Foundation of China.

\newpage
\parindent 0pc
{\large References}\newline
[1] C. R. Hagen, Ann. Phys. {\bf 157}(1984)342.\newline
[2] J. Hong, Y. Kim, and P. Y. Pac, Phys. Rev. Lett.
{\bf 64}(1990)2230;\newline
R. Jackiw and E. J. Weinberg, Phys. Rev. Lett. {\bf 64}(1990)2234.
\newline
[3] R. Jackiw, K. Lee, and E. J. Weinberg,
Phys Rev. D {\bf 42}(1990)3488.\newline
[4] R. Jackiw and S.-Y. Pi, Phys. Rev. Lett. {\bf 64}
(1990)2969; Phys. Rev. D {\bf 42}(1990)3500.\newline
[5] Z. F. Ezawa, M. Hotta, and A. Iwazaki,
Phys. Rev. Lett. {\bf 67}(1991)411; {\bf 67}(1991)\newline
1475(E); R. Jackiw and S.-Y. Pi, Phys. Rev. Lett.
{\bf 67}(1991)415.\newline
[6] R. Jackiw and S.-Y. Pi,Phys. Rev. D {\bf 44}(1991)2524;\newline
Q.-G. Lin, Phys. Rev. D {\bf 48}(1993)1852.\newline
[7] G. W. Semenoff, Phys. Rev. Lett. {\bf 61}(1988)517;
{\bf 63}(1989)1026;\newline
C. R. Hagen, Phys. Rev. Lett. {\bf 63}(1989)1025.\newline
[8] R. Banerjee, Phys. Rev. Lett. {\bf 69}(1992)17;
{\bf 70}(1993)3519;\newline
C. R. Hagen, Phys. Rev. Lett. {\bf 70}(1993)3518.\newline
[9] Q.-G. Lin, Commun. Theor. Phys., to appear.\newline
[10] P. A. M. Dirac, {\it Lectures on Quantum Mechanics}
(Yeshiva University, New York, 1964).\newline
[11] D. Boyanovsky, E. T. Newman, and C. Rovelli,
Phys. Rev. D {\bf 45}(1992)1210.\newline
[12]L.Faddeev and R.  Jackiw, Phys. Rev. Lett. {\bf 60}(1988)1692.
\newline
[13] S. N. Gupta, {\it Quantum Electrodynamics} (Gordon and
Breach, New York, 1977).\newline
\end{document}